\documentclass[twocolumn,prb,floatfix,showpacs]{revtex4}
\usepackage{graphicx}

\begin{document}
%\draft

\title{Magnetic momentum density, Fermi surface and \\
directional magnetic Compton profiles
in LaSr$_{2}$Mn$_{2}$O$_{7}$ and La$_{1.2}$Sr$_{1.8}$Mn$_2$O$_7$
}

\author{P.E. Mijnarends$^{1,2}$, S. Kaprzyk$^{1,3}$, B.Barbiellini$^1$,
Yinwan Li$^{4,5}$, J.F. Mitchell$^5$, P.A. Montano$^{4,6}$, and
A. Bansil$^1$}

\affiliation{$^1$Physics Department, Northeastern University, Boston MA
02115 \\
$^2$Department of Radiation, Radionuclides \& Reactors,
Faculty of Applied Sciences,    \\
Delft University of Technology, Delft, The Netherlands \\
$^3$Academy of Mining and Metallurgy AGH, 30059 Krakow, Poland  \\
$^4$Department of Physics, University of Illinois, Chicago IL 60680 \\
$^5$Materials Science Division, Argonne National Laboratory, Argonne IL
60439 \\
$^6$Scientific User Facilities Division, U.S. Department of Energy, 1000
Independence Avenue, Washington DC 20585-1290}

\date{\today}

\begin{abstract}

We have carried out first principles, all-electron computations of the
magnetic momentum density $\rho_{mag}({\mathbf p})$ and magnetic Compton
profiles (MCPs) for momentum transfer along the [100], [001], and [110]
directions in LaSr$_2$Mn$_2$O$_7$ and La$_{1.2}$Sr$_{1.8}$Mn$_2$O$_7$
within the local spin density approximation (LSDA) based band theory
framework. Parallel measurements of these three MCPs from a single crystal
of La$_{1.2}$Sr$_{1.8}$Mn$_2$O$_7$ at 5 K in a magnetic field of 7 T are
also reported. $\rho_{mag}({\mathbf p})$ is shown to contain distinct
peaks arising from the occupied majority-spin $t_{2g}$ electrons and to
display images of the Fermi surface (FS) in the first and higher Brillouin
zones (BZs). The overall shape of the MCPs, $J_{mag}(p_z)$, obtained by
integrating $\rho_{mag}({\mathbf p})$ over $p_x$ and $p_y$, is found to be
dominated by the majority-spin $t_{2g}$ states. The FS-related fine
structure in the MCPs is however substantial only in the [100] MCP, which
contains features arising from the large majority-spin hole sheets. The
overall shapes and widths of the experimental MCPs along all three
directions investigated are in reasonably good accord with theoretical
predictions, although some discrepancies indicating inadequacy of the LSDA
in treating the magnetic states can be identified. We discuss details of
the FS-related signatures in the first and higher BZs in the [100] MCP and
show that high resolution magnetic Compton scattering experiments with a
momentum resolution of 0.1 a.u. FWHM (full-width-at-half-maximum) or
better will be necessary to observe this fine structure. We comment also
on the feasibility of using positron annihilation spectroscopy in this
connection.

\end{abstract}


\maketitle

\section{Introduction}
\label{sec:intro}

Layered manganites have drawn a great deal of attention as model systems
which display a wide range of electric and magnetic properties and undergo
a variety of phase transitions related to the interplay between the
charge, spin and orbital degrees of freedom as a function of temperature
and doping. The recent revival of interest in the manganites is partly due
to their structural resemblence to the cuprates. The structure is given by
the general formula (La$_{1-x}$Sr$_x$MnO$_3$)$_{n}$SrO (LSMO) and consists
of $n$ layers of corner sharing MnO$_6$ octahedra, separated by
insulating (La,Sr)O layers. The manganites show various degrees of
colossal magnetoresistance (CMR) effect. For example, in the $n=2$ bilayer
material La$_{1.2}$Sr$_{1.8}$Mn$_2$O$_7$ (i.e., $x = 0.40$) investigated
in this study, the CMR effect is a factor of $\sim 200$ at 129 K (just
above the Curie temperature $T_c \sim 120$ K) under a magnetic field of 7
T, and even at low fields the resistance changes by $\sim 200
\%$.\cite{moritomo96} Neutron studies\cite{hirota98} in this compound show
that below $T_c$ the spins are aligned ferromagnetically within a MnO
plane and antiferromagnetically canted between two planes within a double
layer.

Compton scattering refers to inelastic x-ray scattering in the
deeply inelastic regime and it is well known that this technique
provides a unique probe of the correlated many-body ground state
wavefunction of the electronic system via a measurement of the
associated momentum density. In comparison to other {\bf
k}-resolved spectroscopies, Compton scattering possesses the
advantages that it is neither surface sensitive like
photoemission, nor defect sensitive like de Haas-van Alphen or
positron annihilation. The possibility of using magnetic Compton
scattering (MCS) to determine the momentum distribution of
magnetic electrons was recognized quite early,\cite{platzman65,sakai76}
but the scattering cross-section in the magnetic channel is typically
several orders of magnitude smaller than for charge scattering.
For this reason, MCS experiments have become practical only in
the last few years with the availability of high-energy,
circularly polarized, x-rays at the synchrotron light sources.

This article presents all-electron, first-principles
computations of the magnetic Compton profiles (MCPs) in
LaSr$_{2}$Mn$_{2}$O$_{7}$ and La$_{1.2}$Sr$_{1.8}$Mn$_{2}$O$_{7}$ within
the LSDA-based band theory framework. Theoretical predictions for a
scattering vector along the three high-symmetry directions [100], [001],
and [110], are compared and contrasted with corresponding measurements of
the MCPs taken from a La$_{1.2}$Sr$_{1.8}$Mn$_{2}$O$_{7}$ single crystal
at 5 K under a magnetic field of 7 T. Our study provides a benchmark for
assessing electron correlation effects beyond the conventional LSDA
framework on the magnetic momentum density, following up on earlier work
on many non-magnetic materials.\cite{sakurai95} We delineate the nature of
the Fermi surface (FS) generated structure in the magnetic momentum density
and
discuss how this structure is reflected in the MCPs for various directions
of the x-ray scattering vector. All existing MCP data on La-manganite --
including the MCP data presented in this article -- have been taken at a
momentum resolution of around 0.4 a.u. FWHM. We will show however that
this level of resolution is not sufficient for investigating FS signatures
in the MCPs and that for this purpose higher resolution experiments are
needed. In this connection, applicability of the positron annihilation
spectroscopy is also considered. These results will serve to guide
future efforts aimed at exploring FS signatures in La-manganite via
high-resolution MCP and positron annihilation experiments.

Concerning relevant previous magnetic Compton scattering studies
on La-manganite, Li {\em et al.}\cite{li04} have focused on the
[110] MCP in La$_{1.2}$Sr$_{1.8}$Mn$_2$O$_7$ and shown that the
shape of the [110] MCP contains a remarkable signature of the
occupancy of the $d_{x^2-y^2}$ electrons. By using a high magnetic
field of 7 T to maintain an electronically homogeneous phase,
these authors demonstrate that changes with temperature in the
number of $d_{x^2-y^2}$ orbitals can be deduced from the [110] MCPs
measured at different temperatures. Koizumi {\em et
al.}\cite{koizumi01,koizumi06} have investigated the doping dependence of
the [100] and [001] MCPs in La-manganite under a relatively low
magnetic field of 2.5 T. They analyze their MCPs in terms of
atomic and cluster type computations of the momentum density to
gain insight into the occupation of magnetic orbitals in the
system but do not obtain detailed FS information.

By way of brief background, we note that the magnetic momentum density
$\rho_{mag}({\mathbf p})$ is given by
\begin{equation}
   \rho_{mag}({\mathbf p}) = \rho_{\uparrow}({\mathbf p}) -
                             \rho_{\downarrow}({\mathbf p}),
\label{eq1}
\end{equation}
where \cite{footnote-2}
\begin{equation}
                \rho_{\uparrow(\downarrow)}({\mathbf p}) =
      \sum_{i}n_i^{\uparrow(\downarrow)}
        |\int \psi_i^{\uparrow(\downarrow)}({\mathbf r})
              \exp(-i{\mathbf p} \cdot {\mathbf r}) d{\mathbf r}|^2
\label{eq2}
\end{equation}
is the spin-polarized momentum density of a system of electrons
in states $i$ with wavefunctions $\psi_i$ and occupation numbers
$n_i$. The Fourier transforms of the wavefunctions generally vary
smoothly with momentum. The occupation function, on the other
hand, introduces discontinuities in $\rho_\uparrow$ and
$\rho_\downarrow$ at the Fermi momenta and at the Umklapp images
of the FS at higher momenta with appropriate weights.

The quantity of interest in a magnetic Compton scattering
experiment is the double differential magnetic scattering
cross-section, given by \cite{OUP04,sakai84}
\begin{eqnarray}
\label{eq2.5}
  \frac{d^2\sigma_m}{d\Omega d\omega_2}& = &
    \frac{r_e^2}{2}\left(\frac{\omega_2}{\omega_1}\right)^2 P_c
     \frac{\omega_1}{m_ec^2} (\cos \theta-1) \nonumber \\
&&[\cos \alpha \cos \theta + (\frac{\omega_2}{\omega_1}) \cos (\alpha -
\theta)]
J_{mag}(p_z),
\end{eqnarray}
where $P_c$ is the degree of circular polarization, $\theta$ is
the scattering angle, and $\omega_1(\omega_2)$ is the photon
energy before (after) scattering. $\alpha$ is the angle between
the spin (assumed to lie in the scattering plane) and the
momentum of the incident photon, $m_e$ is the electron rest mass,
and $r_e$ is the classical electron radius. Finally, the MCP
$J_{mag}(p_z)$ is defined as the difference between the Compton
profiles for the majority ($\uparrow$) and minority
($\downarrow$) spin profiles $J_\uparrow$ and $J_\downarrow$:
\begin{eqnarray}
 J_{mag}(p_z) &=&  J_{\uparrow}(p_z) - J_{\downarrow}(p_z)\nonumber \\
              &=&   \int \int [\rho_{\uparrow}({\mathbf p})
                  - \rho_{\downarrow}({\mathbf p})] dp_x dp_y.
\label{eq3}
\end{eqnarray}
It should be noted that $J_{mag}(p_z)$ only involves unpaired
electron spins. Unlike neutron scattering, the magnetic Compton
experiment does not couple to the orbital moment.

The remainder of this article is organized as follows.
Section~\ref{sec:exp}
describes experimental details. In
Sec.~\ref{sec:comp} we discuss earlier band-structure studies on
the 327 and 113-manganites, followed by particulars of our electronic
structure calculations based on the Korringa-Kohn-Rostoker (KKR)
method to obtain spin-polarized energy bands, Fermi surfaces,
momentum densities, and the MCPs. 
Section~\ref{subsec:BS} briefly presents
band-structure results, while Sec.~\ref{subsec:FS} considers
the Fermi surface, the way it changes under doping, and the
possible role it plays in the occurrence of structural and
magnetic instabilities. The spin-dependent momentum density
and the MCPs are
presented in Sec.~\ref{subsec:MCP}. Section \ref{subsec:posi}
discusses in how far positron annihilation may be able to
shed further light on details of the Fermi surface. The article
concludes with a summary of the results in Sec.~\ref{sec:summ}.

\section{Experiments}
\label{sec:exp}

Circularly polarized photons were produced using an Elliptical
Multipole Wiggler (EMW) at beamline 11-ID-B (BESSRC) of the
Advanced Photon Source at Argonne National Laboratory.
The horizontal field in the EMW is produced by
electromagnets so that the degree of polarization $P_c$
can be tuned by adjusting the current through the
magnets.\cite{montano95} This allows optimization of the
figure of merit $P_c \sqrt{I}$ (where $I$ is the intensity
of the beam) and easy reversal of the
polarization, important when working at high magnetic
fields where fast switching of the field is difficult.
The Laue monochromator consisted of an annealed Si(220)
crystal, 10mm wide and 10 mm thick, mounted on a water-cooled
Ni-plated Cu support in Ga. In addition, a water-cooled Cu
filter was mounted in front of the crystal to reduce the
heat load.
The sample was mounted inside the cryostat and could be
magnetized by a field of up to 8 T generated by an Oxford
Instruments Spectromag superconducting magnet. The
temperature at the sample position could be varied from
1.7 to 300 K. The sample was a high-quality single crystal
of La$_{1.2}$Sr$_{1.8}$Mn$_2$O$_7$ measuring
$10 \times 5 \times 2$ mm$^3$ with its shortest side along
the crystalline $c$ axis. It was fixed to the holder by a
high-purity aluminum clip to avoid magnetic contamination.
The x-ray beam intersects neither the holder nor the clip.
A motorized slit was placed in the beam close to the sample.
Its size was chosen to fit the sample dimensions in
order to maximize the figure of merit.
The scattering angle $\theta$ was set at 170$^{\circ}$,
while the magnetic field was oriented at an angle of
$\alpha = 176^{\circ}$ with respect to the incident beam.
The scattered photons were detected by a three-element
Ge solid-state detector with a resolution of 0.4 keV at
100 keV. All measurements reported here were performed at
125 keV. Magnetic Compton profiles were obtained using magnetic
fields of 3 T and 7 T along the three high-symmetry directions
[001], [100], and [110] at temperatures of 5, 100, 160, and 200 K,
which straddle the Curie temperature $T_c = 120$ K. In this article
only the 5 K, 7 T measurements will be discussed as these are
magnetically homogeneous and therefore the most relevant. The momentum
resolution was 0.4 a.u. FWHM.  \\

\section{Computations}
\label{sec:comp}

\subsection{Earlier band-structure studies}
\label{subsec:earlier}

There exist only a few studies of the electronic structure of the
bilayer manganites. De Boer and de Groot\cite{deboer99,footnote-1} performed 
a full-potential LAPW calculation of the electronic structure of
LaSr$_2$Mn$_2$O$_7$ using the structural data of Seshadri {\em et
al.}\cite{seshadri97} Exchange-correlation effects were incorporated
within the framework of the generalized gradient approximation
(GGA).\cite{perdew92} These authors obtained a metallic majority-spin band
while the minority-spin band is reminiscent of a doped semiconductor with
a band gap of 1.7 eV. The Fermi level lies slightly above the bottom of
the minority-spin conduction band, yielding a nearly half-metallic
ferromagnet. The total spin magnetic moment per formula unit is found to
be 6.995 $\mu_B$. (If the Fermi level were lying in the gap the magnetic
moment would be 7 $\mu_B$.) Huang {\em et al.}\cite{huang00} calculated
the electronic structure of La$_{2-2x}$Sr$_{1+2x}$Mn$_2$O$_7$ for $x$
between\cite{footnote0} 0.3 and 0.5 for different lattice
parameters,\cite{kubota00}
using the full-potential linear muffin-tin orbital (LMTO)
method\cite{methfessel88,methfessel89} with the Ceperly-Alder
exchange-correlation functional.\cite{ceperley80} They used the virtual
crystal approximation (VCA) to treat the effect of
La/Sr substitution and also found the material to be (nearly) half-metallic
for $x$ between 0.4 and 0.5. These authors also computed the Fermi surface
which was found to be strongly two-dimensional. The question of whether
La$_{1.2}$Sr$_{1.8}$Mn$_{2}$O$_{7}$ is a half-metallic ferromagnet or not
is still a controversial one. As noted above, the band structure of Ref.
\onlinecite{deboer99} shows small minority-spin pockets at $\Gamma$,
whereas they seem to be absent in Ref.~\onlinecite{huang00}. Recent
angle-resolved photoemission studies\cite{dessau06} may directly support
their existence.

More work has been done on the related perovskite
manganites of general formula La$_{1-x}$(Ca,Sr)$_{x}$MnO$_3$,
which also show the CMR effect. Pickett and Singh\cite{pickett96,singh98}
have shown that an LSDA calculation for the
undoped end-compound LaMnO$_3$ using a 5 atom perovskite unit
cell produces a ferromagnetic (FM) ground state. However, if a 20
atom $\sqrt 2 \times 2 \times \sqrt 2$ {\em Pnma} supercell is
used and the structure is allowed to relax (resulting in
rotations of the O octahedra and Jahn-Teller (JT) distortions), a
small band gap opens, in combination with new band splittings and
shifts. As a result, the ground state is found to be an A-type
antiferromagnetic (AFM) insulator (spin-aligned in layers,
alternating from layer to layer\cite{wollan55}) in agreement with
experiment. CaMnO$_3$, on the other hand, is correctly found to
be a G-type AFM insulator with a narrow band gap and a rocksalt
arrangement of moments. Thus, the LSDA produces the right ground
state in both cases.

As LaMnO$_3$ is hole-doped by partial substitution of
La$^{3+}$ by a divalent element (Ca, Sr, Ba) ($x \approx 0.33$, the
region where CMR occurs), the ground state becomes ferromagnetic with
rotated octahedra but no JT distortion.\cite{singh98} Since the
Mn $d$ band is now less than half occupied the Fermi level $E_F$ is
not close to a (pseudo) band gap for the majority-spin electrons
and the material is a metal. Livesay {\em et al.}\cite{livesay99}
have calculated the electronic structure of
La$_{0.7}$Sr$_{0.3}$MnO$_3$ and performed positron annihilation
measurements of the Fermi surface. They find a FS
consisting of hole cuboids centered at R (coined 'woolsacks' by
the authors) and an electron spheroid centered at $\Gamma$, which
touches the woolsacks along the [111] directions. This FS
can be viewed as a three-dimensional analogue of the
two-dimensional FS found in
La$_{2-2x}$Sr$_{1+2x}$Mn$_2$O$_7$ (cf. also Sec.~\ref{subsec:FS}).
Similar to its 2D equivalent, the presence of flat parts of the FS
opens the possibility of nesting. Other similarities may be found
in the density of states. All these calculations agree on the 
presence of small minority-spin electron pockets, similar to those
resulting from electronic structure calculations on 327-LSMO.
Also, point-contact Andreev reflection measurements\cite{nadgorny01}
on La$_{0.7}$Sr$_{0.3}$MnO$_3$ have shown the presence of minority-spin
electrons at the FS.
Finally, we performed LSDA calculations of the FM
phases of both LaMnO$_3$ and LaSr$_2$Mn$_2$O$_7$. The projected
densities of states for these FM cases look very similar. For the
majority-spin channel the Mn $3d$ states of both compounds form
bands between -2.5 eV and +2.5 eV with respect to $E_F$. The
$e_g$ bands are rather broad compared to the $t_{2g}$ bands,
cross $E_F$, and are therefore only partly filled. The exchange
interaction places them ~2.5 eV higher in energy than the
majority-spin states. In summary, important differences between
the two classes of materials notwithstanding, there are also
strong parallels to be found.

\subsection{Present work}

In order to obtain the Fermi surface, momentum densities and Compton
profiles, the electronic structure of LaSr$_2$Mn$_2$O$_7$ was first
obtained within an all-electron fully charge and spin self-consistent KKR
framework.\cite{kaprzyk90,bansil91,bansil92,bansil99,footnote0.5} The
formalism for computing momentum densities is discussed in Refs.
\onlinecite{mijnarends76,mijnarends79,mijnarends90,bansil81,mijnarends95}.
The
structure data were taken from Seshadri {\em et al.}\cite{seshadri97} for
the space group I4/mmm (No.~139).\cite{footnote0.75} Two empty spheres per
formula unit were inserted to increase the filling factor. The maximum
angular momentum cut-off {\em l}$_{max}$ was 3 for all
atoms.\cite{footnote0.85} Exchange-correlation effects were incorporated
within the von Barth-Hedin local spin density approximation
(LSDA).\cite{lsda} The self-consistency cycles were repeated until the
maximum difference between the input and output potentials was less than
$10^{-5}$ Ry. Parallel to this, we have also carried out full-potential
computations within both the KKR and the LAPW schemes in order to
ascertain that the band structure and Fermi surface underlying our
momentum density and MCP computations is essentially the same as the
full-potential results. Moreover, we have carried out self-consistent
computations on La$_{1.2}$Sr$_{1.8}$Mn$_2$O$_7$
within the VCA scheme to study
the effect of La/Sr substitution, and find that
the rigid band approximation is appropriate.
However, there might be some uncertainty
with regard to where the
doped magnetic electrons should go.
We have modeled the effect by simply
adding the extra 0.2 electrons
per formula unit exclusively into the majority spin band,
which is similar to the viewpoint
argued by Huang {\em et al.}\cite{huang00}
and in keeping with the spirit of Hund's rules.
In principle, one could imagine that
some of the added electrons will
in fact get added into the minority
spin bands and
in that case the effect of doping on
the majority sheets will be reduced.

The spin dependent momentum density was computed on a regular mesh in {\bf
p} space, created by translating 765 uniformly spaced {\bf k} points in
the 1/16-th irreducible wedge of the Brillouin zone via 36000 reciprocal
lattice vectors {\bf G} each (i.e., Umklapp processes {\bf p} = {\bf k} +
{\bf G}). Thus, the resulting mesh size is $\Delta p_x = \Delta p_y
=0.02681$ a.u. and $\Delta p_z =0.04157$ a.u., with the mesh filling a
sphere of radius 14.9 a.u. This dataset of band-by-band
$\rho_{\uparrow}({\mathbf p})$ and $\rho_{\downarrow}({\mathbf p})$ values
over the aforementioned mesh in {\bf p} space forms the basis for
obtaining the directional MCPs and 2D-projections for the stoichiometric
compound. By moving the Fermi level for the majority-spin electrons up by
20 mRy to account for the extra electrons, the same dataset is used to
repeat the computations for the doped compound.

\section{Results and discussion}
\label{sec:resdis}

\subsection{Band structure}
\label{subsec:BS}

There is a good overall agreement between our band structure
and     the results of Refs.~\onlinecite{deboer99} and
\onlinecite{huang00}. All three computations agree on a nearly
halfmetallic ferromagnetic band structure with the Fermi
level crossing Mn $d$ bands.
In all computations the majority-spin bands
are metallic with a Fermi surface consisting of three sheets.
An analysis of the wavefunctions indicates that the majority-spin
bands at $E_F$ mainly involve a mixture of La and Mn $e_g$, and
Sr and O $s-p$ states, although the O $2p$ and Mn $t_{2g}$
character increases rapidly as one moves to higher binding
energies. On the whole, there is little dispersion in the $c$
direction, so that the electronic structure is largely
two-dimensional.
The minority conduction band, which near $\Gamma$ dips below
$E_F$, consists of two nearly degenerate bands (0.7 mRy apart at
$\Gamma$) of antibonding Mn $d_{xy}$ character. Overall, the
minority-spin band structure is similar to that of a doped n-type
semiconductor with $E_F$ intersecting the bottom of the
conduction band.\cite{footnote1.5}
\begin{figure}
\begin{center}
\includegraphics[width=\hsize]{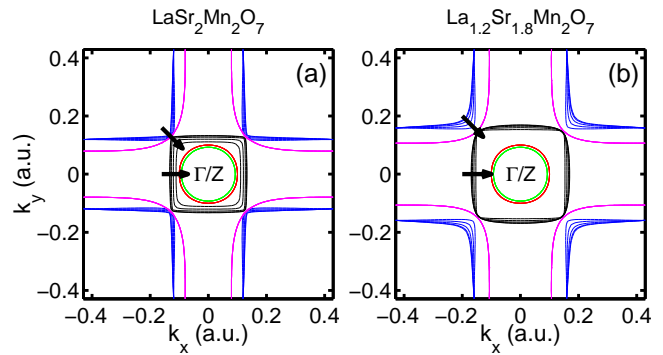}
\end{center}
\caption{(color online)
Fermi surface of (a) LaSr$_{2}$Mn$_{2}$O$_{7}$ and
(b)La$_{1.2}$Sr$_{1.8}$Mn$_{2}$O$_{7}$. For each FS sheet five
intersections with $(00k_z)$ planes are shown for $k_z$ at $\Gamma$, $k_z$
at Z=$(0,0,2\pi/c)$, and $k_z$ at three points between $\Gamma$ and Z.
Some intersections
are indistinguishable from each other due to the lack of $k_z$-dispersion.
Arrows indicate how the FS cross section changes with increasing $k_z$.
Blue and red lines give the two majority spin hole sheets; the squarish
feature in the center (black lines) is the majority spin electron pillar.
The two circular FS sheets at the center (red and green lines) are the
minority-spin electron pillars.
}
\label{fig1}
\end{figure}

\subsection{Fermi surface}
\label{subsec:FS}

Figure 1 shows cross sections of the Fermi surface at five different $k_z$
values for the stoichiometric compound LaSr$_{2}$Mn$_{2}$O$_{7}$ ($x=0.5$)
as well as the doped material La$_{1.2}$Sr$_{1.8}$Mn$_{2}$O$_{7}$
($x=0.4$).\cite{footnote1.6} Although the two sets of FSs are similar in
shape, doping by substituting La for Sr increases the electron count and
thus changes FS dimensions. Both FSs show three sheets for the
majority-spin electrons, which consist of two large hole cylinders
centered around the X ($\pi/a,\pi/a$)-points (red and blue contours), and
one squarish $\Gamma$-centered electron pillar (black
contours).\cite{footnote0.85} The arrows indicate how the contours change
with increasing $k_z$. For example, the $\Gamma$-centered majority-spin
electron pillar (black lines) in (a) decreases in size with increasing
$k_z$. One set of majority spin hole cylinders has a squarish
cross-section with small $k_z$ dispersion near the corners (blue lines),
while the other set (red lines) displays strongly rounded corners and no
noticeable $k_z$ dispersion. The two small $\Gamma$-centered circles in (a)
and (b) depict the two dispersionless minority-spin sheets, which have the
form of circular pillars.

The flatness of some of the FS sheets of undoped LSMO gives rise to
Fermi-surface nesting, which can drive structural or magnetic
instablities. For instance, the flat parts of the X-centered hole
cylinders span nesting vectors {\bf q} between $\sim (0.18,0)2\pi/a$ and
$(0.28,0)2\pi/a$, and have been implicated in giving rise to an
incommensurate charge density wave in the $x$ and $y$ directions, coupled
with Jahn-Teller distortions.\cite{campbell01,chuang01}
With doping [Fig.~1(b)],
both these hole cylinders become smaller and the splitting between the two
hole cylinders increases. There also are changes in the shapes of the
various FS sheets with doping (e.g., the blue hole cylinder becomes more
rounded at corners in the $k_z$=0 plane and the black electron cylinders
become more bulged on the sides), and     associated changes in the
nesting properties of the FS.

The flat nature of parts of the Fermi surface, together with its
pronounced two-dimensional character, also make LSMO an obvious candidate
for observation of the FS via the MCS technique. Since according to Eq.
(\ref{eq2}) the momentum density is not only determined by the occupation
function but also by the wavefunction transforms, we investigate in the
following section how the interplay between FS and wavefunctions affects
the observability of various FS sheets. The complete Fermi surface can in
principle be mapped without the interference of wavefunction effects by
performing a 2D- or 3D-reconstruction of the momentum density followed by
Lock-Crisp-West folding as described by Matsumoto {\em et
al.}\cite{matsumoto01} for Al-3at.\% Li alloys. In this way, one can
transform the momentum density in {\bf p} space to an occupation number
density in {\bf k} space.

\subsection{Magnetic Compton profiles}
\label{subsec:MCP}

The MCPs are obtained from the 3D magnetic momentum density
$\rho_{mag}({\mathbf p})$ via Eq.~(\ref{eq3}). It is useful, however, to
take an intermediate step and first consider 2D-projections along various
high-symmetry directions by performing only one of the two integrations
involved in Eq.~(\ref{eq3}). Such a 2D distribution can be obtained by
reconstructing $\rho_{mag}({\mathbf p})$ from a series of measured MCPs
and, excepting for effects of the non-uniform positron spatial
distribution, it is also measured in a 2D angular correlation of positron
annihilation radiation (2D-ACAR) experiment.\cite{west95} A
$(\gamma,e\gamma)$ experiment where the kinematics of the scattered photon
is measured in coincidence with that of the outgoing recoil electron can
also access this 2D-distribution in
principle.\cite{schneider92,bell91,tschentscher93}

\begin{figure}
\begin{center}
\includegraphics[width=\hsize]{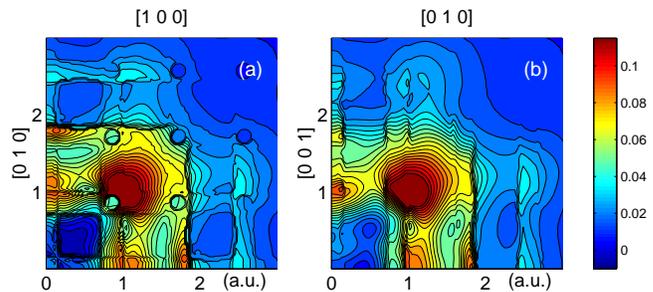}
\end{center}
\caption{(color online)
2D-projections of the magnetic momentum density
$\rho_{mag}(\bf p)$ for the doped material
La$_{1.2}$Sr$_{1.8}$Mn$_{2}$O$_{7}$ onto (a) the
(001) plane  and (b) the (100) plane.
}
\label{fig2}
\end{figure}

\begin{figure}
\begin{center}
\includegraphics[width=\hsize]{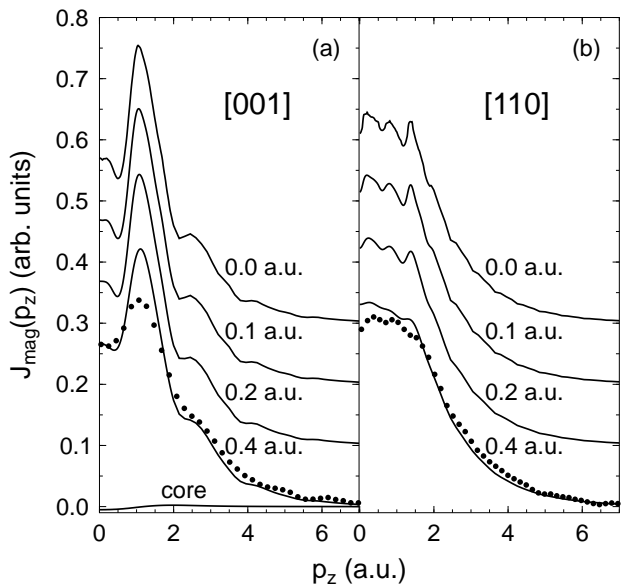}
\end{center}
\caption{
Computed magnetic Compton profiles for the [001] and [110] directions in
La$_{1.2}$Sr$_{1.8}$Mn$_{2}$O$_{7}$ convoluted with different resolution
functions of indicated FWHM. The profiles include a small, isotropic core
contribution shown separately at the bottom of (a). The dots give the
experimental profiles at 7 T. The size of the dots is representative of the
error bars on the data points. All profiles are normalized to the same
area and have been offset vertically for clarity.
}
\label{fig3}
\end{figure}

\begin{figure}
\begin{center}
\includegraphics[width=\hsize]{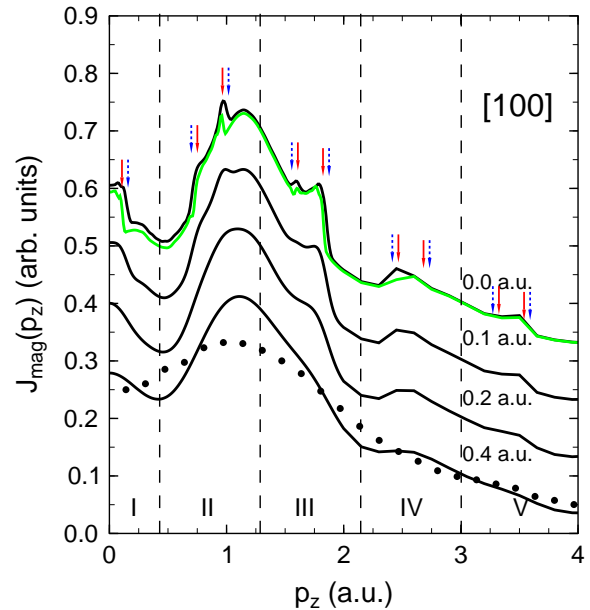}
\end{center}
\caption{
(color online) Central part of the computed magnetic Compton profiles
for the [100] direction in LaSr$_{2}$Mn$_{2}$O$_{7}$ (green line) and
La$_{1.2}$Sr$_{1.8}$Mn$_{2}$O$_{7}$ (black lines) convoluted with
different resolution functions of indicated FWHM. The red (solid)
and blue (dashed) arrows
indicate the projected positions of the flat red and blue pieces of the
doped Fermi surface in Fig.~1(b). Vertical dashed lines denote boundaries
between the Brillouin zones numbered by the Roman numerals at the bottom.
}
\label{fig4}
\end{figure}

Figure~\ref{fig2} shows 2D-projections of $\rho_{mag}({\mathbf p})$ in the
doped material onto the (001) and (100) planes. Note that Fermi breaks
occur in the momentum density in the first as well as higher BZs, the
latter due to Umklapp processes. The size of each break depends on the
matrix element of the band intersecting the Fermi surface. Some Fermi
breaks are therefore clearly seen while others are hardly visible. In
general the breaks are largest in the 2nd and 3rd BZ as a consequence of
the predominant $d$ character of the electrons at the Fermi energy in the
manganite.\cite{footnote2.5} We now consider the (001) projection in
Fig.~\ref{fig2}(a)
and focus on signatures therein of various FS sheets of Fig. 1(b).
Some of the flat faces of the large majority spin hole-cylinders and their
Umklapp images in higher BZs are seen clearly. The splitting between the
two types of large hole cylinders ($\sim 0.04$ a.u.) is difficult to see
on the scale of the figure. With regard to the $\Gamma$-centered squarish
majority-spin electron pillar, two sides of this pillar are visible near
the little circular feature located at (0.858, 0.858) a.u.
This feature is one of a system of small circular features located on a
square mesh of interval 0.858 a.u. They          arise from the small
minority-spin electron pillars at $\Gamma$ (cf.~Fig.~1) and are
particularly noticeable along the [110] direction and its vicinity,
consistent with the $d_{xy}$ symmetry of the
underlying wavefunctions.\cite{footnote2.7,harthoorn78} In
the following we shall see that the minority-spin pillars are too small to
survive the second integration in Eq.~(\ref{eq3}) and do not produce a
measurable signature in the MCPs. The question of their observability
through positron annihilation spectroscopy is considered in
Sec.~\ref{subsec:posi} below. The results of Figure~\ref{fig2}(b), which
gives the (100) projection of the magnetic momentum density, can be
interpreted along lines very similar to the preceding discussion of the
(001) projection. The presence of clear vertical features in
Fig.~\ref{fig2}(b)
again emphasizes the 2D character of the FS.
The [001], [110], and [100] MCPs, shown in Figs.~\ref{fig3} and \ref{fig4},
involve a further integration of the 2D-projections of Fig.~\ref{fig2}
along a high-symmetry direction. For example, the [100] MCP in
Fig.~\ref{fig4}
is obtained by projecting the 2D-distribution of Fig.~\ref{fig2}(a) onto
the
[100] (horizontal) axis, i.e., by integrating this density along the [010]
direction. The MCPs in Figs.~\ref{fig3} and \ref{fig4} include a small,
isotropic
contribution due to the polarization of the occupied core orbitals, which
is given separately at the bottom of Fig.~\ref{fig3}(a). We see at once
that the high-density region in Fig.~\ref{fig2}(a) centered around (1
a.u.,1 a.u.) projects into the peak extending over 0.5-1.8 a.u. in
the [100] MCP in Fig.~\ref{fig4}. This is also the case with the [001] MCP
in
Fig.~\ref{fig3}(a), where one integrates the density of Fig.~\ref{fig2}(b)
over the horizontal direction. In sharp contrast, the [110] MCP requires
an integration along a [1$\overline{1}$0] direction in Fig.~\ref{fig2}(a),
which causes a broad peak centered
around $p=0$ and spread over 0-1.8 a.u. in the [110] MCP in
Fig.~\ref{fig3}(b).\cite{li04,footnote3}

In discussing baseline FS signatures in the MCPs it is best to focus on
the unbroadened theory curves, which are the topmost curves in
Figs.~\ref{fig3} and \ref{fig4}; we return to comment on the important role
of
resolution below. We first observe that due to the different projections
of the magnetic momentum density involved in various profiles, the [100]
profile is the only one expected to show significant FS structure. We
therefore turn to Fig.~\ref{fig4}, which shows the [100] MCP on an
expanded scale. As in Fig.~\ref{fig3}, the black solid curves in
Fig.~\ref{fig4}
refer to La$_{1.2}$Sr$_{1.8}$Mn$_{2}$O$_{7}$. The green curve represents
the unbroadened MCP for the undoped ($x=0.5$) compound. The figure shows
that the effect of doping, corresponding to the addition of 0.2
electron/f.u., is to raise the MCP by a small amount to account for the
extra added electrons. Beyond this, there are shifts in the FS related
features, which are in accord with the changes in the FS dimensions
discussed in connection with Fig. 1 above. The blue and red arrows give
the positions at which the blue and red flat pieces of the Fermi surface
(the large hole cylinders) in Fig.~1(b) project onto the [100] direction.
The projected sides of the majority-spin electron pillar nearly coincide
with those of the smaller of the two hole cylinders and are not shown
separately.
The vertical dashed lines show the BZ boundaries. The arrows in zone I
indicate the FS breaks in the first BZ; all other arrows correspond to
Umklapp processes in higher BZs. The hole cylinders should thus manifest
themselves as depressed parts of the profile between the two neighboring
pairs of arrows surrounding various BZ boundaries. This is seen indeed to
be the case between the arrows around the boundary of zones I and II over
0.2-0.7 a.u. as well as between zones III and IV over 1.85-2.40 a.u. This
depression is however less clear between zones II and III or between zones
IV and V due to compensating effects of other peaks in the momentum
density such as the peak from the filled $t_{2g}$ states. These results
further highlight the importance of the matrix element in shaping FS
features. For example, the size of the break at 1.85 a.u. is much larger
than at 1 a.u. or at 2.5 a.u.

We now turn to the question of resolution broadening. Figures
\ref{fig3} and \ref{fig4} show how the computed MCPs and the FS
signatures therein are affected by resolution broadening varying
from 0 - 0.4 a.u. (FWHM) and how these theoretical predictions
compare with the corresponding experimental results. Recalling
that the resolution broadening in the present measurements is 0.4
a.u., we see that the computed and measured MCPs are in
reasonably good overall accord along all three high-symmetry
directions. The computations correctly reproduce the presence of
the strong peak around 1.1 a.u.~in the [100] and [001] MCPs in
Figs.~\ref{fig4} and \ref{fig3}(a) as well as the rather flat
behavior of the [110] MCP at low momenta in Fig.~\ref{fig3}(b).
Some of the theoretically predicted secondary structure is also
recognizable in the measurements [e.g., the bump around 2.5 a.u.
in Fig.~\ref{fig3}(a)], but under the resolution of 0.4 a.u. most
FS features in the MCPs are washed out as are other smaller
structures (e.g., the three small peaks in Fig.~\ref{fig3}(b) at
low momenta).

There are discrepancies as well between theory and experiment in
Figs.~\ref{fig3} and \ref{fig4}. Notably, the computed peak in
the [100] and [001]
MCPs around 1.1 a.u. is stronger than the measured one. This difference
reflects the limitations of the present LDA framework in treating
correlation effects on the magnetic electrons in the
system.\cite{footnote3.5,lam74,kaplan03} Electron correlations are known
to broaden structure in Compton profiles and move momentum density from
low to higher momenta.\cite{sakurai95} They may also
reduce the sizes of Fermi breaks in the momentum density in comparison to
LSDA predictions. It is therefore important to establish whether the
experimental results are resolution limited or not. A look at the
different resolution broadened curves in Figs.~\ref{fig3} and \ref{fig4}
indicates that in
order to obtain insight into this issue and to pin down FS features in the
MCPs a resolution of 0.1 a.u. or better will be necessary.

\subsection{Positron annihilation}
\label{subsec:posi}

\begin{figure}
\begin{center}
\includegraphics[width=\hsize]{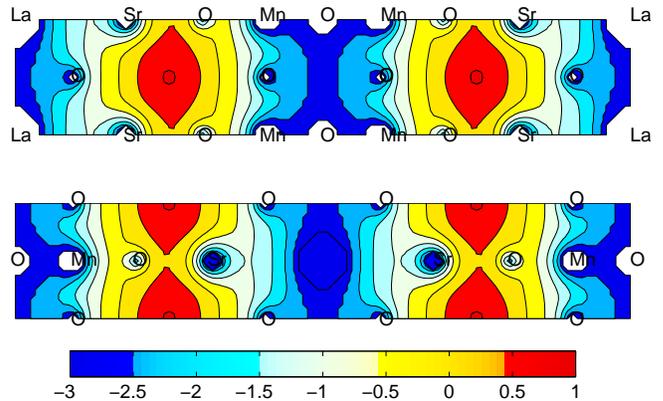}
\end{center}
\caption{(color online)
Logarithmic plot of the positron density in the planes $y=0$ (top) and
$y=b/2$ (bottom) of the unit cell of LaSr$_2$Mn$_2$O$_7$. The $c$-axis of
the cell is drawn horizontally.
}
\label{fig5}
\end{figure}

In view of the similarity between the 2D-projection of $\rho_{mag}(\mathbf
p)$ in Fig.~\ref{fig2} and a polarized positron (e$^+$) 2D-ACAR
distribution the interesting question arises whether the minority-spin
pockets might
be visible in a polarized positron measurement.\cite{hanssen90}
                                                The 2D-ACAR technique
avoids the second integration involved in Eq. 4 and possesses a superior
momentum resolution (0.03-0.08 a.u.), enough to detect the minority-spin
pillars, which have a diameter of 0.18 a.u. On the other hand, the
positron does not sample the unit cell uniformly, and therefore, the
positron wavefunction would need to overlap the Mn $d_{xy}$ orbitals
responsible for these pillars, in order to observe these pillars via the
positron techniqe. A computation of the e$^+$ density shown in
Fig.~\ref{fig5}        finds that the positron does not entirely avoid the
Mn
$d_{xy}$ orbitals, even though the greatest positron density is found
between the SrO layers.          A computation of the magnetic (001)
2D-ACAR distribution (not shown here), however, shows that the positron
overlap with these Mn orbitals is indeed too low and that the
minority-spin electron pillars are therefore not observable in a polarized
e$^+$ 2D-ACAR experiment. The majority-spin electron cylinders and pillars
on the other hand are observable.
A polarized e$^+$ 2D-ACAR experiment would be interesting in this
connection.

\section{Summary and conclusions}
\label{sec:summ}

We present all-electron first-principles computations of the
magnetic momentum density $\rho_{mag}({\mathbf p})$ and MCPs along three
high-symmetry directions in LaSr$_2$Mn$_2$O$_7$ and
La$_{1.2}$Sr$_{1.8}$Mn$_2$O$_7$ within the LSDA-based band theory
framework. Parallel measurements of these three MCPs from a single crystal
of La$_{1.2}$Sr$_{1.8}$Mn$_2$O$_7$ at 5 K under a magnetic field of 7 T are
also reported. The band structure is found to be that of a nearly
half-metallic ferromagnet with a small minority-spin FS. The
spin-polarized densities $\rho_\uparrow$ and $\rho_\downarrow$ are
relatively smooth and structureless and are dominated by the large number
of occupied bands. Their difference, $\rho_{mag}({\mathbf p})$, however,
contains clear features due to the filled majority-spin $t_{2g}$ states,
the minority-spin counterparts of which lie above $E_F$. Insofar as FS
signatures are concerned, imprints on the theoretical MCPs of the large
squarish majority-spin hole sheets as well as those of the small
minority-spin electron pillars are clearly seen spread throughout momentum
space, although the intensity is generally small in the low momentum
region due to the non-$s$ character of the associated wavefunctions.

The quantity measured in a magnetic Compton scattering experiment is not
$\rho_{mag}({\mathbf p})$, but the MCP $J_{mag}(p_z)$, which is given by a
double integral over $p_x$ and $p_y$ via Eq. (\ref{eq3}). It is therefore
important to understand how various magnetic orbitals and FS sheets will
be reflected in the MCPs. We discuss this point in some detail. It turns
out that the filled $t_{2g}$ states of $xy$, $yz$, and $zx$ symmetry
continue to dominate the overall shape of the MCPs. These are responsible
for the pronounced peaks around 1.1 a.u. in the [100] and the [001] MCP
and also for the rather flat behavior of the [110] MCP over the 0 - 1.8
a.u. momentum range. Out of the five FS sheets in our band structure, only
the large majority-spin hole cylinders yield a substantial fine structure
in the computed [100] MCP; the [001] and [110] directions of the
scattering vector are not favorable for investigating FS signatures in the
MCP. The majority as well as the minority-spin electron pillars are barely
visible in the MCPs and their observation through the magnetic Compton
scattering technique will require reconstruction of 2D or 3D magnetic
momentum density from high resolution MCPs taken along a series of
directions. The polarized positron annihilation 2D-ACAR experiment is also
            sensitive in observing some of these FS features.

The overall shapes and widths of the experimental MCPs along all three
directions investigated are in good accord with theoretical predictions.
In particular, the measured [100] and [001] MCPs display a pronounced peak
around 1.1 a.u., while the [110] MCP is quite flat at low momenta, as is
the case in the computed MCPs. A notable discrepancy is that the height of
the measured peak around 1.1 a.u. \protect in the [100] and [001] MCPs is
smaller than the computed one, suggesting that electron correlation
effects beyond the LSDA are needed to fully describe the magnetic
electrons.\cite{footnote5} Some of the theoretically predicted fine
structure is recognizable in the measured MCPs, but under the experimental
resolution of 0.4 a.u., much of this structure including FS features is
essentially washed out. Our analysis provides the momentum regions that
deserve close scrutiny in order to detect Fermi surface related structure.
It shows that a momentum resolution of about 0.1 a.u. or better is needed
to pin down these FS features and other fine structure in the MCPs. It is
clear that further magnetic Compton scattering
measurements as well as polarized e$^+$ 2D-ACAR measurements on
La-manganites would be of great interest. \\

\acknowledgments

We thank Hsin Lin for technical assistance 
in connection with this study. 
This work was supported by the US Department of Energy
contracts DE-AC03-76SF00098 and DE-FG02-07ER46352.
It was also sponsored by the
Stichting Nationale Computer Faciliteiten for the use of
supercomputer facilities, with financial support from the
Netherlands Organization for Scientific Research (NWO), and
benefited from the allocation of supercomputer time at the NERSC
and the Northeastern University's Advanced Scientific Computation
Center (NU-ASCC).

\begin{noindent}

\end{noindent}

\begin{thebibliography}{99}
\bibitem{footnote-2}
 Assuming an independent-particle model in which the occupation 
 function is one for an occupied state and zero for an unoccupied state.
\bibitem{footnote0}
  Note that our definition of $x$ is different
  from that of Ref. \onlinecite{huang00}.
\bibitem{footnote0.5}
  In our computations the core states are treated
  relativistically, but the valence electrons are treated
  semi-relativistically.
\bibitem{footnote2.7}
  Since the projected magnetic momentum density in the little circles is
  smaller than the surrounding density, the contribution of the
  minority-spin pillars clearly outweighs that of the majority-spin
  pillars.
\bibitem{footnote3}
  The increased difference at low momenta between theory and experiment
  with respect to Fig. 2 of Ref.~\onlinecite{li04} is the result of doping.
\bibitem{footnote3.5}
  It should be noted that the present computation does not include
  a Lam-Platzman (Ref. \onlinecite{lam74})
  correction to take into account electron-electron correlation. It is not
  clear how the Lam-Platzman type correction can be generalized to treat
  the magnetic momentum density.
\end{thebibliography}
\end{document}